\documentclass[11pt]{article}
\pdfoutput=1
\usepackage{amsmath,amssymb,amsthm,amsxtra,overpic,bbm,bm,epsfig,slashed}
\usepackage{color,subfigure}
\usepackage{hyperref}
\usepackage[numbers,sort&compress]{natbib}
\usepackage{ulem}
\def\thefootnote{\fnsymbol{footnote}}
\allowdisplaybreaks[4]

\def\thefootnote{\fnsymbol{footnote}}

\newcommand{\bq}{\begin{eqnarray}}
\newcommand{\nq}{\end{eqnarray}}

\newcommand{\ini}{{\bf i}}
\newcommand{\fin}{{\bf f}}

\def\dx{\mathrm{d}x}
\def\dy{\mathrm{d}y}
\def\dz{\mathrm{d}z}

\def\mnui{m_{\bf i}}
\def\mnuj{m_{\bf f}}

\begin{document}

\begin{center}
{\Large\bf $CP$ asymmetries in the rare top decays $t\to c\gamma$ and $t\to c g$ }
\end{center} 
\vspace{0.2cm}

\begin{center}
{\bf Shyam Balaji}\footnote{Email: \tt shyam.balaji@sydney.edu.au} ~
\\\vspace{5mm}
{School of Physics, The  University of Sydney, NSW 2006, Australia} \\

\end{center}

\vspace{1.5cm} 

\begin{abstract} 
The rare radiative flavour changing top decays $t\to c\gamma$ and $t\to cg$ (and the even rarer $t\to u\gamma$ and $t\to ug$) have been processes of interest for decades as they offer a key probe for studying top quark properties. However an explicit analytical study of the branching ratios and $CP$ asymmetries resulting from these loop level processes has thus far evaded attention. In this work, we provide the formulation for the $CP$ asymmetry resulting from the total kinetic contribution of the loop integrals and their imaginary parts, as well as an updated numerical computation of the predicted Standard Model (SM) branching fractions. These rare processes are suppressed in the SM by the Glashow-Iliopoulos-Maiani (GIM) mechanism. However, the results presented here can easily be exported for use in minimal extensions of the SM including vector-like quarks or in Two-Higgs-Doublet models where radiative fermionic decay processes can be enhanced relative to the SM by several orders of magnitude. Such processes provide an experimentally clean signature for new fundamental physics and can potentially be tested by current collider experiments. These topical beyond the SM theories are an elegant means to provide improved global fits to the latest results emerging from flavour physics, CKM and precision electroweak measurements.
\end{abstract}
\begin{flushleft}
\hspace{0.8cm} PACS number(s): \\
\hspace{0.8cm} Keywords: top quark, $CP$ violation, radiative decay, flavour physics
\end{flushleft}

\def\thefootnote{\arabic{footnote}}
\setcounter{footnote}{0}

\newpage

\section{Introduction}
The study of radiative decays has been of interest for many decades because they provide an experimentally clean probe for new physics \cite{Beneke:2000hk}. The electromagnetic dipole moment of heavy quarks can be generated at various loop levels and their radiative decays are induced by the off diagonal parts of the dipole moments analogously to the lepton sector \cite{Balaji:2019fxd, Balaji:2020fxd}. Precision measurements of electromagnetic interactions provide a tantalising probe for new physics beyond the Standard Model (SM) \cite{Balaji:2020fxd}. This is particularly relevant due to the presence of current top factories such as the Large Hadron Collider (LHC) which provide an unprecedented increase in top quark statistics, thereby enabling radical improvement in the understanding of heavy quark properties \cite{Beneke:2000hk}. Of particular importance are precision studies of the various rare top quark decays. These include flavour-changing neutral (FCN) decays $t\rightarrow c Z$ as well as $t\rightarrow c\gamma$ and $t\rightarrow c g$ \cite{AguilarSaavedra:2002ns}. The radiative decays of heavy fermions are more significant than those of light fermions due to their larger partial widths resulting from their much higher relative mass. Hence, such clean channels are of major importance in testing precise theoretical predictions for particle properties and searching for tensions with the SM. 

Within the SM, these processes are mediated at lowest order in perturbation theory by penguin diagrams with charged down-type quarks running loops. However, due to the large hierarchy in the down-type quark masses relative to the $W$ bosons in the loop, these decays are suppressed by the Glashow-Iliopoulos-Maiani (GIM) mechanism. This is in contrast with processes such as $b \rightarrow s\gamma$, which contain the much heavier top quark in the loop. This extra suppression resulted in branching ratios being computed at $\lesssim 10^{-10}$ or smaller \cite{Eilam:1990zc,Atwood:1996vj,Grzadkowski:1990sm,Luke:1993cy}. These were later estimated with more precision in Ref. \cite{AguilarSaavedra:2002ns}, using the the $b$-quark running mass at the top mass scale in the $\overline{\text{MS}}$ scheme. The use of the running $b$-quark mass represents a more rigorous treatment for the calculation as the top quark decays at its pole mass. 

In this work, we focus primarily on a precise computation of the SM branching ratios for the radiative top decays with the current Cabibbo-Kobayashi-Maskawa (CKM) best fit values and particle masses extracted from Ref. \cite{Tanabashi:2018oca}. Additionally, we pay particular interest to the computation of the $CP$ asymmetry resulting from the imaginary part of the loop integrals that imply $\Gamma(t \rightarrow c\gamma) \neq \Gamma(\bar{t}\rightarrow \bar{c}\gamma)$. We provide the closed form analytical formulation for the kinetic loop terms and their imaginary parts that generate the $CP$ asymmetry. Here we note that by kinetic loop term, we refer to the contribution coming explicitly from the particles running in the loop and not the vertex contributions which can be factorized separately. We will continue with this nomenclature for the rest of this work. This is in contrast to previous studies which are limited to numerical estimations of the loop functions derived from generic Passarino-Veltman functions \cite{AguilarSaavedra:2002ns}. Although in the SM, the radiative process branching ratios are currently unobservable due to the aforementioned large GIM suppression, the above results can be easily applied to a host of beyond the SM theories which we briefly outline below. 

A notable application of the formulation shown could be beyond the SM extensions with heavy vector-like quarks (VLQs) \cite{AguilarSaavedra:2002ns,AguilarSaavedra:2002kr} e.g. heavy $t'$ and $b'$ states with extended Cabibbo-Kobayashi-Maskawa (CKM) matrices. Many of which provide an improved global fit to data compared to the SM when considering several flavour physics observables and precision electroweak measurements \cite{Alok:2015iha,Alok:2015vvk,Alok:2014yua}. A comprehensive review of the various types of VLQs can be found in Ref. \cite{Aguilar-Saavedra:2013qpa} and there is some related discussion in Ref. \cite{Boehm:2003hm}. The addition of quark singlets to the SM particle content represents the simplest way to break the GIM mechanism and can thereby enable large radiative decay widths. These models typically contain a non-unitary higher dimensional CKM matrix and contain flavour changing neutral couplings (FCNC) to the $Z$ boson at tree-level since the new heavy quarks are not $SU(2)_L$ doublets.

Moreover, there are other SM extensions that can enhance branching ratios for top decays by many orders of magnitude thereby yielding compelling phenomenology. For instance, in Two-Higgs-Doublet models (2HDM) we find that $\mathcal{B}(t \rightarrow cZ) \sim 10^{-6}$, $\mathcal{B}(t \rightarrow c\gamma) \sim 10^{-7}$, $\mathcal{B}(t \rightarrow cg) \sim 10^{-5}$ can be achieved \cite{Atwood:1996vj}. More recently, it was shown that in the type-III 2HDM one could expect up to $N(t\rightarrow c\gamma)=100$ events at the LHC with an integrated luminosity of $300fb^{-1}$ in certain parameter regions \cite{Gaitan:2015hga}. The rare top quark decays at one-loop with FCNCs coming from additional fermions and gauge bosons has been studied in several extensions of the SM such as the minimal super-symmetric model, Left-Right symmetry models, top colour assisted technicolour and two Higgs doublets with four generations of quarks \cite{Dedes:2014asa, Hill:1994hp, Gaitan:2015hga, Atwood:1996vj,Arhrib:2005nx,DiazCruz:1989ub,Atwood:1995ud, Atwood:1995ej}.
There is also potential for similar radiative processes to occur in models with leptoquarks such as light versions of the ones shown in \cite{Balaji:2018zna,Balaji:2019kwe}.

These applications are of particular interest, since it was recently shown that a net circular polarisation, specifically an asymmetry between two circularly polarised photons $\gamma_{+}$ and $\gamma_{-}$, is generated if $CP$ is violated in neutrino radiative decays \cite{Boehm:2017nrl}. The same $CP$ effect is induced for top quarks or new VLQs and therefore polarisation measurements on the resulting photons are a crucial and experimentally clean probe for new physics.

The outline of the paper is as follows, we first show the full radiative process calculation in Section \ref{section:radiativecalculation}. This section is further divided into an overview of the interaction Lagrangian, computation of the relevant amplitudes, analytical evaluation of the kinetic terms and most importantly their imaginary parts (which are responsible for generating $CP$ asymmetry), followed by showing the computation for the $CP$ asymmetry itself. This is accompanied by Section \ref{section:results} which contains an overview of the process to calculate the radiative branching fractions and decay widths for the various channels as well as the main numerical results. Finally, we briefly discuss the applications of the formalism to beyond the SM theories in Section \ref{section:BSMmodels} via inclusion of heavy VLQs and the 2HDM.

\section{Calculation of radiative processes}
\label{section:radiativecalculation}
\subsection{Calculation of Lorentz invariant amplitudes}
\label{section:amplitudes}
In this work we first overview the interaction Lagrangian relating the mass eigenstates of the up and down-type quarks via the SM charge current interaction. We denote the up-type quarks as $u_\beta = (u,c,t)$ and the down-type quarks as $d_\alpha = (d,s,b)$. The corresponding interaction Lagrangian is then given by
\begin{equation}
\mathcal{L}_{int}  = -\frac{g}{\sqrt{2}}\left[\bar{u}_\beta\gamma^\mu P_L  V_{\beta\alpha}d_\alpha\right]W_\mu^{+} + h.c .
\label{eq:Lagrangian}
\end{equation}
Where $V$ is the SM $3\times3$ CKM matrix and $g$ is the usual weak interaction gauge coupling constant and $ P_L $ is the left-chiral projection operator.

\begin{figure}[ht!]
\centering
    \includegraphics[width=0.9\linewidth]{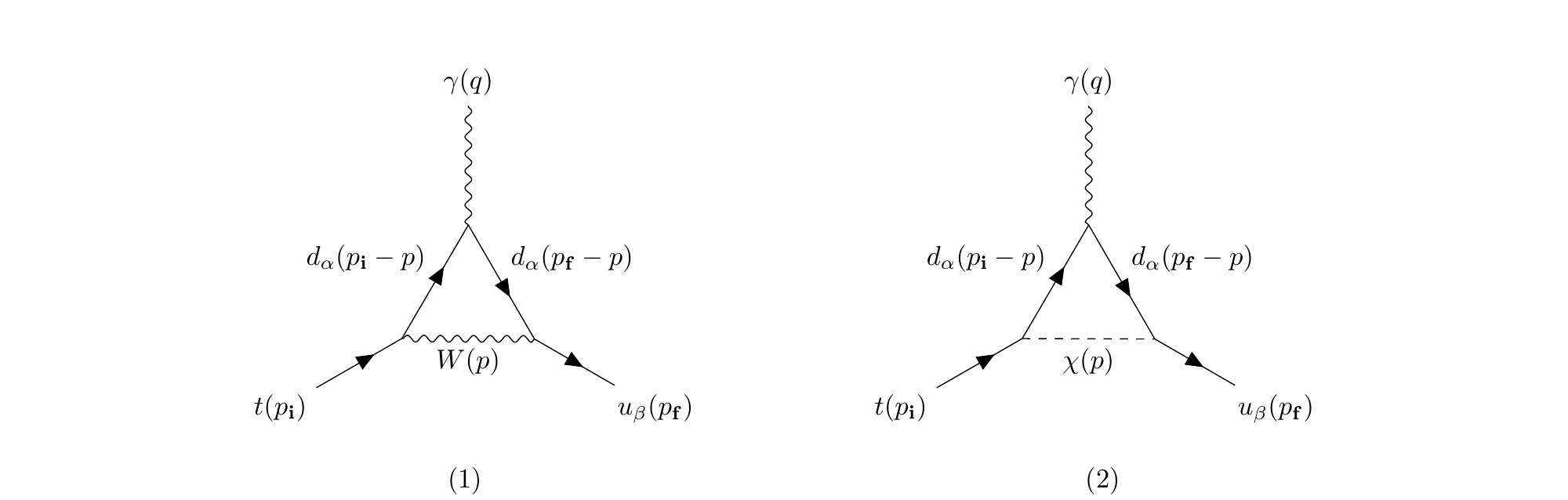}
    \includegraphics[width=0.9\linewidth]{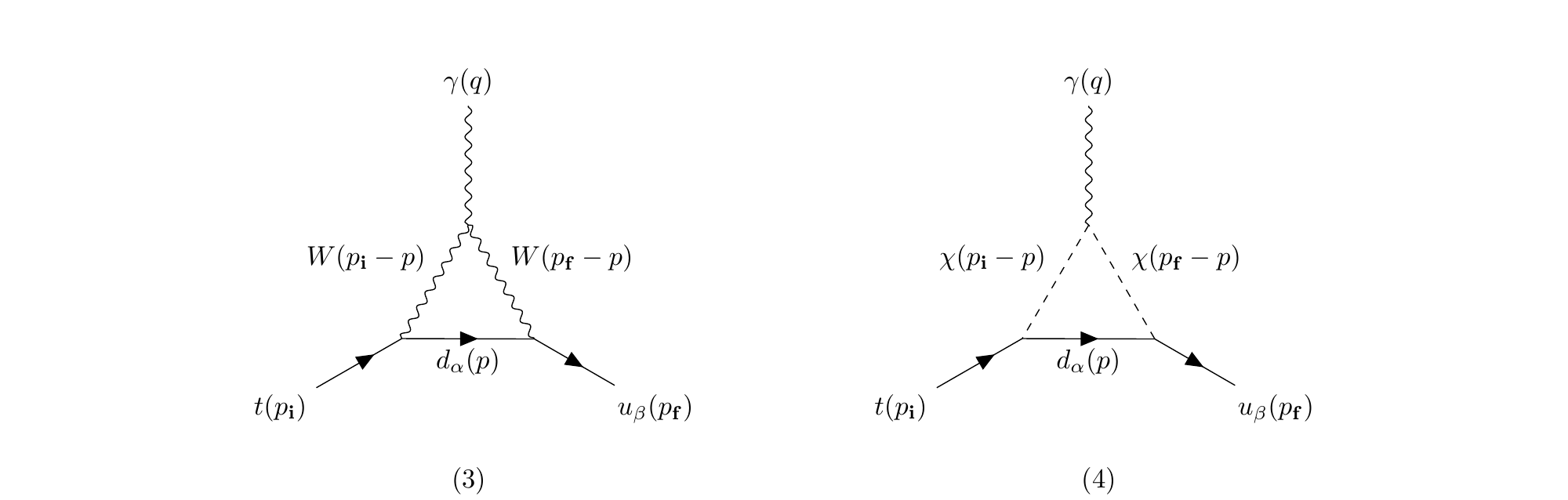}
    \includegraphics[width=0.9\linewidth]{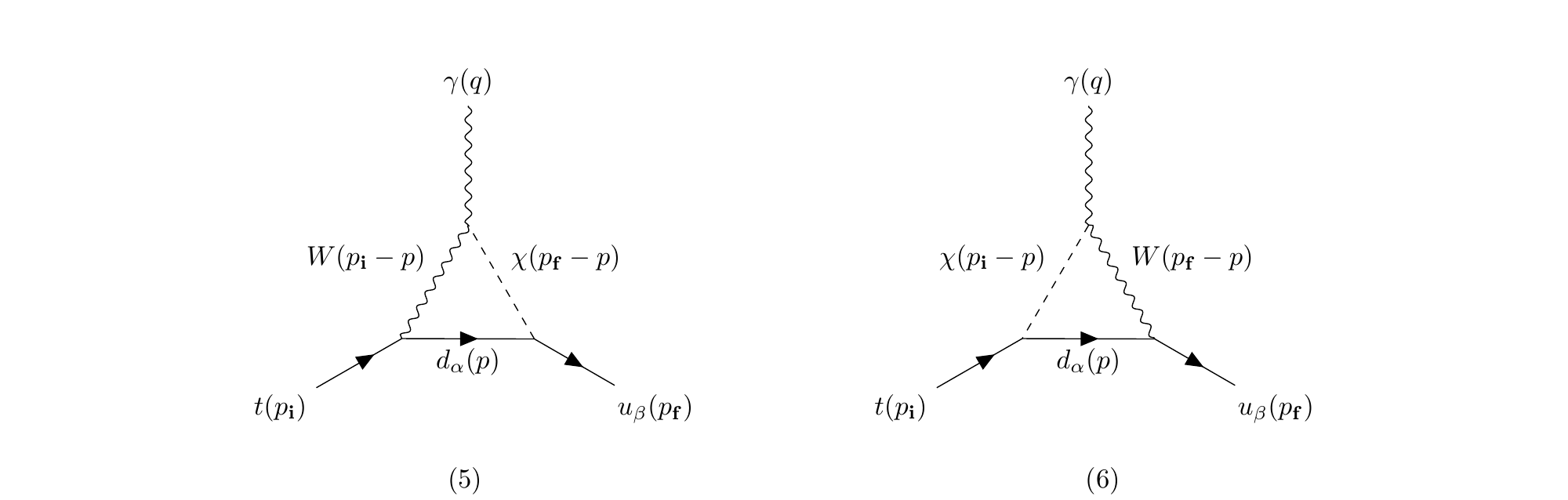}
    \caption{Feynman Diagrams for the one-loop radiative top decay $t\rightarrow u_\beta \gamma$ induced by SM weak interactions with SM fields. We denote the amplitudes for the six diagrams as $\mathcal{M}_1-\mathcal{M}_6$ accordingly. For the gluon channel, $t\rightarrow u_\beta g$, only the first two diagrams contribute and the photon is replaced with a gluon. In all cases, external radiated gauge boson momenta is denoted $q=p_\fin-p_\ini$ while the internal momenta that are integrated over in the loop calculations are denoted $p$. }\label{figure:FeynmanDiagramsgamma}
\end{figure}

We focus firstly on the contributions to the rare photon radiative top decay mediated by SM interactions as given in Figure~\ref{figure:FeynmanDiagramsgamma}. In this work, we are primarily interested in top decays, hence we denote the initial state $t$ and the final state quark to be generically $u_\beta=(u,c)$ and $d_{\alpha}=(d,s,b)$. Hence we may write the corresponding $t\rightarrow u_\beta \gamma$ process amplitudes in full generality as follows 
\begin{eqnarray}
\!\!\! \! i \mathcal{M} (t \to u_\beta + \gamma_{\pm})  = i \bar{u}(p_\fin) \Gamma_{\fin \ini }^\mu(q^2) u(p_\ini) \varepsilon^*_{\pm,\mu}(q) \,. \label{eq:decay_amplitude}
\end{eqnarray}
More explicitly, for each Feynman diagram shown in Figure~\ref{figure:FeynmanDiagramsgamma}, we have
\begin{align}
 \!\! \! \! i\mathcal{M}_1&= i{\frac{e g^{2}}{6}}  V_{t \alpha } V^*_{\beta\alpha }\!\! \int\!\! \frac{d^4p}{(2\pi)^4}\frac{\overline{u}(p_\fin) \gamma_\mu P_L (\slashed{p}_\fin -\slashed{p}+m_d)\gamma^\rho(\slashed{p}_\ini-\slashed{p}+m_d)\gamma^\mu P_L  u(p_\ini) \epsilon^*_\rho(q)}{[(p_\fin-p)^2-m_d^2][p^2-m_W^2][(p_\ini-p)^2-m_d^2]},\nonumber\\[.2cm]
\! \!\!\!i\mathcal{M}_2&=  i{\frac{eg^{2}}{6m_W^2}}V_{t \alpha } V^*_{\beta\alpha } \!\! \int\!\! \frac{d^4p}{(2\pi)^4}\frac{\overline{u}(p_\fin)(m_\beta P_L -m_d
     P_R )(\slashed{p}_\fin-\slashed{p}+m_d)\gamma^\rho(\slashed{p}_\ini-\slashed{p}+m_d)(m_d P_L -\mnui P_R )u(p_\ini) \epsilon^*_\rho(q)}{[(p_\fin-p)^2-m_d^2][(p_\ini-p)^2-m_d^2][p^2-m_W^2]}\nonumber,\\[.2cm]
 \!\! \! \!i\mathcal{M}_3&= i{\frac{eg^{2}}{2}}   V_{t \alpha } V^*_{\beta\alpha } \!\! \int\!\! \frac{d^4p}{(2\pi)^4}\frac{\overline{u}(p_\fin) \gamma_\nu P_L  (\slashed{p}+m_d)\gamma_\mu P_L  V(p_\ini,p_\fin,p)^{\mu\nu\rho} u(p_\ini) \epsilon^*_\rho(q)}{[(p_\fin-p)^2-m_W^2][p^2-m_d^2][(p_\ini-p)^2-m_W^2]}\nonumber,\\[.2cm]
   \!\! \! \! i \mathcal{M}_4&= i{\frac{eg^{2}}{2m_W^2}} V_{t \alpha } V^*_{\beta\alpha } \!\! \int\!\! \frac{d^4p}{(2\pi)^4}\frac{\overline{u}(p_\fin)(\mnuj P_L -m_d
     P_R )(\slashed{p}+m_d)(m_d P_L -\mnui P_R )(2p-p_\ini-p_\fin)^\rho u(p_\ini) \epsilon^*_\rho(q)}{[(p_\fin-p)^2-m_W^2][p^2-m_d^2][(p_\ini-p)^2-m_W^2]}\nonumber,\\[.2cm]
   i  \mathcal{M}_{5+6}&= i{\frac{eg^{2}}{2}}   V_{t \alpha } V^*_{\beta\alpha } \!\! \int\!\! \frac{d^4p}{(2\pi)^4}\overline{u}(p_\fin)\left[\frac{\gamma^\rho P_L  (\slashed{p}+m_d)(m_d P_L -\mnui P_R )}{(p^2-m_d^2)((p_\fin-p)^2-m_W^2)((p_\ini-p)^2-m_W^2)}\right.\nonumber\\[.2cm]
      &\ \,\,\,\,\,\,\,\,\,\,\,\,\,\,\ \,\,\,\,\,\,\,\,\,\,\,\,\,\,\ \,\,\,\,\,\,\,\,\,\,\,\,\,\,\ \,\,\,\,\,\,\,\,\,\,\,\,\,\,\ \,\,\,\,\,\,\,\,\,\,\,\,\,\, \left.-\frac{(m_\beta P_L -m_d P_R )
    (\slashed{p}+m_d)\gamma^\rho P_L  }{(p^2-m_d^2)((p_\ini-p)^2-m_W^2)((p_\fin-p)^2-m_W^2)}\right]u(p_\ini) \epsilon^*_\rho(q).\label{eq:loop1-6}
\end{align}
where the contribution from the triple gauge boson vertex is given
\begin{eqnarray}
V^{\mu\nu\rho} &=& g^{\mu\nu}(2p_\ini-p-p_\fin)^\rho +
  g^{\rho\mu}(2p_\fin-p-p_\ini)^\nu +
  g^{\nu\rho}(2p-p_\ini-p_\fin)^\mu \,,
\end{eqnarray}

and $e$ refers to the usual $U(1)$ Abelian electromagnetic charge. We denote the initial state momentum of the top quark as $p_\ini$ and the final state up-type quark as $p_\fin$. The 't Hooft-Feynman gauge is chosen to simplify the amplitude calculations and the scalar $\chi$ refers to the unphysical charged Goldstone boson. We apply the  Gordon decomposition as well as Ward identity $q_\mu \mathcal{M}^\mu = 0$ and ignore all vector terms proportional to $\gamma^\mu$, since these are simply vertex corrections to the overall electric charge, we need only consider tensor-like terms within the current $\Gamma_\mu$ to determine the transition form factor resulting from these diagrams. 

We follow the standard procedure to integrate over all internal momenta  $p$ in the loop with the help of the Feynman parametrisation. We take the initial and final state chiralities into account followed by factorising the electromagnetic dipole moment terms with coefficients as 
\begin{eqnarray}\label{eq:effective_vertex0}
&\Gamma_{\fin \ini, \alpha}^{\mu,({\rm k})}&= \frac{e g^2}{4(4\pi)^2} V_{ \ini\alpha} V^*_{ \fin \alpha}
i \sigma^{\mu\nu}q_\nu \int_0^1\dx\dy\dz\,\delta(x+y+z-1)\, {\cal P}^{(\rm k)} \,.
\end{eqnarray}
where each loop contribution is given by
\begin{eqnarray}\label{eq:effective_vertex1}
&{\cal P}^{(1)}&= \frac{-2x (x+z) m_\ini P_\text{R} - 2x (x+y) m_\fin P_\text{L}}
{3 \Delta_{\alpha W}(x,y,z)}\,, \nonumber\\
&{\cal P}^{(2)}&= \frac{[x z m_\fin^2 -((1 - x)^2 +xz) m_d^2] m_\ini P_\text{R} + [xy m_\ini^2 - ((1-x)^2 + xy) m_d^2] m_\fin P_\text{L}}
{3 m_W^2 \Delta_{\alpha W}(x,y,z)}\,, \nonumber\\
&{\cal P}^{(3)}&= \frac{[(1-2x)z-2(1-x)^2] m_\ini P_\text{R} + [(1-2x)y-2(1-x)^2] m_\fin P_\text{L}}
{\Delta_{W \alpha}(x,y,z)}\,, \nonumber\\
&{\cal P}^{(4)}&= \frac{[xz m_\fin^2 - x (x+z) m_d^2] m_\ini P_\text{R} + [xy m_\ini^2 - x (x+y) m_d^2] m_\fin P_\text{L}}
{m_W^2 \Delta_{W \alpha}(x,y,z)}\,, \nonumber\\
&{\cal P}^{(5)}&= \frac{-z m_\ini P_\text{R}} {\Delta_{W \alpha}(x,y,z)}\,, \nonumber\\
&{\cal P}^{(6)}&= \frac{-y m_\fin P_\text{L}} {\Delta_{W \alpha}(x,y,z)}\,, 
\end{eqnarray}
 where it is convenient to define the function in the denominator in terms of the Feynman parameters as 
 \begin{eqnarray} \label{eq:Deltaxyz}
\Delta_{W \alpha}(x,y,z) &=& m_W^2(1-x)+x m_d^2-x (y m_\ini^2 + z m_\fin^2), \nonumber\\
\Delta_{\alpha W}(x,y,z) &=& m_d^2(1-x)+x m_W^2-x (y m_\ini^2 + z m_\fin^2) \,. 
\end{eqnarray}   
We note that for $t\rightarrow u_\beta g$, the structure of  amplitudes are largely the same, but we only require $3{\cal P}^{(1)}$ and ${3\cal P}^{(2)}$ (because the down-quark electric charge prefactor of $Q=\frac{1}{3}$ doesn't appear at the highest vertex) along with the gauge coupling replacement $e\rightarrow g_s$ in Eq.~\eqref{eq:effective_vertex0} due to the presence of gluon emission.

\subsection{Derivation of the total kinetic contribution}
\label{section:kineticterms}
We are now ready to compute the total kinetic contribution for both $t\rightarrow u_\beta \gamma$ and $t\rightarrow u_\beta g $ channels. From Ref. \cite{Balaji:2019fxd}, it was shown we could rewrite Eqs.~\eqref{eq:effective_vertex0}, \eqref{eq:effective_vertex1} and \eqref{eq:Deltaxyz} in terms of the dimensionless kinetic term ${\cal F^\gamma}$ such that
\begin{eqnarray}\label{eq:effective_vertex}
&\Gamma_{\fin \ini, \alpha}^{\mu,({\rm k})}&= \frac{eG_\text{F}}{4\sqrt{2}\pi^2} V_{\ini\alpha } V^*_{\fin \alpha }
i \sigma^{\mu\nu}q_\nu ({\cal F}_{\fin \ini, \alpha}^\gamma  m_\ini P_\text{R}+ {\cal F}_{\ini \fin, \alpha}^\gamma m_\fin P_\text{L}) \,.
\end{eqnarray}
In the case of a gluon being radiated instead of a photon (which is otherwise identical to the first two diagrams in Figure \ref{figure:FeynmanDiagramsgamma}), we simply make the coupling replacement $e\rightarrow g_s$ in the above expression as well as $\mathcal{F}^\gamma\rightarrow \mathcal{F}^g$. Performing the loop integrals using the same approach shown in Ref. \cite{Balaji:2020fxd} and summing the kinetic contribution for each individual diagram $\sum_{k=1}^{5}{\cal P}^{(k)}$ with a radiated photon results in 
\begin{align}
\label{eq:F_photon_integral}
{\cal F}^\gamma_{\fin \ini, d}&=\int_0^1\dx \left\{\frac{(m_\ini^2-m_d^2-2m_W^2)(m_d^2+m_\fin^2x^2)+x m^4_{\fin\ini,d}}{3(m_\fin^2-m_\ini^2)^2x}\log\left(\frac{m_d^2+x(m_W^2-m_d^2-m_\ini^2)+m_\ini^2x^2}{m_d^2+x(m_W^2-m_d^2-m_\fin^2)+m_\fin^2x^2}\right)\right.\nonumber\\[.5cm]
&\left.+\frac{(m_\ini^2-m_d^2-2m_W^2)(m_d^2+m_\fin^2(x-1)^2)+(1-x)m^4_{\fin\ini,d}}{(m_\ini^2-m_\fin^2)^2x}\log\left(\frac{m_W^2+(m_d^2-m_W^2-m_\ini^2)x+m_\ini^2}{m_W^2+(m_d^2-m_W^2-m_\fin^2)x+m_\fin^2}\right)\right\}\nonumber\\[.5cm]\,&\hspace{2cm}+\frac{2(m_d^2-m_\fin^2+2m_W^2)}{3(m_\fin^2-m_\ini^2)}.
\end{align}
We also consider the case where a gluon is radiated which only corresponds to the first two diagrams i.e. $\sum_{k=1,2} 3{\cal P}^{(k)}$ where, as mentioned earlier, the pre-factor of three is required since the down-quark electric charge $Q=\frac{1}{3}$ does not appear at the quark-quark-gluon vertex, therefore we may write ${\cal F}^{g}$ as  
\begin{align}
\label{eq:F_gluon_integral}
{\cal F}^g_{\fin \ini, d}&=\int_0^1\dx \left\{\frac{(m_\fin^2 - 2 m_W^2 - m_d^2) (x-1) x}{(m_\fin^2-m_\ini^2)x}\right.\nonumber\\[.5cm]
&\,\left.+\frac{(m_\ini^2-m_d^2-2m_W^2)(m_d^2+m_\fin^2x^2)+x m^4_{\fin\ini,d}}{(m_\ini^2-m_\fin^2)^2x}\log\left(\frac{m_d^2+(m_W^2-m_d^2-m_\ini^2)x+m_\ini^2}{m_d^2+(m_W^2-m_d^2-m_\fin^2)x+m_\fin^2}\right)\right\},
\end{align}
where in both cases we make the assignment
\begin{equation}
   m^4_{\fin\ini,d}=2m_W^2m_d^2-(m_d^2+m_\fin^2-2m_W^2)(m_\ini^2-m_d^2-m_W^2).
\end{equation}
We note that in Eq.~\eqref{eq:F_photon_integral} and Eq.~\eqref{eq:F_gluon_integral} the sub-index $d$ denotes each flavour of down-type quark that can run in the loop, this will later have to be summed over when computing branching ratios and $CP$ observables.

The non-zero imaginary parts for ${\cal F}_{\fin \ini, \alpha}^{\gamma,g}$ and ${\cal F}_{\ini \fin, \alpha}^{\gamma,g}$ can now be obtained. 
Since, they include integral terms of the form $\int _0^1 \dx f (x) \log g (x)$, where $g(x)$ is not positive definite in $(0,1)$. One can instead use the fact that there is an interval $(x_1,x_2) \subset (0,1)$ where $g(x)<0$ is satisfied, and $x_1$ and $x_2$ are solutions of $g(x)=0$. The real and imaginary parts in the integration can then be split into 
\begin{eqnarray}
\int _0^1 \dx f (x) \log g (x)=\int _0^1 \dx f(x) \log|g(x)|+ i\pi \int_{x_1}^{x_2} \dx f(x) \,.
\end{eqnarray}
Now the imaginary part given by $\int_{x_1}^{x_2}\dx f (x)$ can be analytical obtained. 
In this way, we derive the following key analytical expressions
\begin{align}
\label{eq:imaginarypartFgamma}
    \text{Im}[{\cal F}_{\fin \ini, d}^\gamma]= &\left\{\frac{\pi \vartheta(m_\ini-m_W-m_d) }{3 (m_\fin^2-m_\ini^2)^2}\left[\frac{\mu_\ini^2}{m_\ini^4}\rho^6+m_d^2 \left(m_\ini^2-m_d^2-2 m_W^2\right) \log \left(\frac{m_\ini^2+m_d^2-m_W^2+\mu_\ini^2}{m_\ini^2+m_d^2-m_W^2-\mu_\ini^2}\right)\nonumber\right.\right.\\
    &\left.\left.-3m_W^2 \left(m_d^2+m_\fin^2-2 m_\ini^2+2 m_W^2\right) \log \left(\frac{m_\ini^2+m_W^2-m_d^2+\mu_\ini^2}{m_\ini^2+m_W^2-m_d^2-\mu_\ini^2}\right)\right]\right\}\nonumber \\
  + &\left\{\frac{\pi \vartheta(m_\fin-m_W-m_d)}{3(m_\fin^2-m_\ini^2)^2}\left[\frac{\mu_\fin^2}{m_\fin^2}\sigma^4-m_d^2(m_\ini^2-m_d^2-2m_W^2)\log\left(\frac{m_\fin^2+m_d^2-m_W^2-\mu_\fin^2}{m_\fin^2+m_d^2-m_W^2+\mu_\fin^2}\right)\right.\right.\nonumber\\
  &\left.\left.+ 3m_W^2 \left(
m_d^2+m_\fin^2-2 m_\ini^2+2 m_W^2\right)\log\left(\frac{m_\fin^2+m_W^2-m_d^2-\mu_\fin^2}{m_\fin^2+m_W^2-m_d^2+\mu_\fin^2}\right)  \right] \right\},
   \end{align}
and similarly
\begin{align}
\label{eq:imaginarypartFg}
    \text{Im}[{\cal F}_{\fin \ini, d}^g]&=\left\{\frac{\pi \vartheta(m_\ini-m_W-m_d)}{2 (m_\fin^2-m_\ini^2)^2}\left[\frac{\mu_\ini^2}{m_\ini^4}\xi^6-2 m_d^2  (m_d^2 - m_\ini^2 + 2 m_W^2)  \log \left(\frac{m_\ini^2+m_d^2-m_W^2+\mu_\ini^2}{m_\ini^2+m_d^2-m_W^2-\mu_\ini^2}\right)\right]\right\}\nonumber \\
  &+\left\{\frac{\pi \vartheta(m_\fin-m_W-m_d)}{2(m_\fin^2-m_\ini^2)^2}\left[\frac{\mu_\fin^2}{m_\fin^2}\eta^4-2 m_d^2  (m_d^2 - m_\ini^2 + 2 m_W^2)\log\left(\frac{m_\fin^2+m_d^2-m_W^2+\mu_\fin^2}{m_\fin^2+m_d^2-m_W^2-\mu_\fin^2}\right) \right] \right\},
   \end{align}
where the following mass dimension parameters $\rho, \sigma, \xi$ and $\eta$ are introduced as
\begin{align}
   \rho^6=& \left(m_d^2-m_\ini^2\right) \left(m_d^2 \left(m_\fin^2-2 m_\ini^2\right)+2 m_\fin^2 m_\ini^2\right)+m_W^2 \left(m_d^2 \left(m_\fin^2-2 m_\ini^2\right)+7 m_\fin^2 m_\ini^2-4 m_\ini^4\right)-2 m_W^4 \left(m_\fin^2-2 m_\ini^2\right),\nonumber\\
   \sigma^4=& 2 m_\fin^2 \left(m_d^2-m_\ini^2+3 m_W^2\right)+m_\ini^2 \left(m_d^2-3 m_W^2\right)+(m_W^2-m_d^2) \left(m_d^2+2 m_W^2\right),\nonumber\\
   \xi^6=&(m_d^2 - 
    m_\ini^2) ( (2 m_d^2 + 
       m_\fin^2) m_\ini^2-m_d^2 m_\fin^2 ) - (m_\fin^2 m_\ini^2 - 4 m_\ini^4 + 
    m_d^2 (m_\fin^2 - 2 m_\ini^2)) m_W^2 + 2 (m_\fin^2 - 2 m_\ini^2) m_W^4,\nonumber\\
    \eta^4 =& (m_d^2 + m_\fin^2) (m_d^2 - m_\ini^2) + (m_d^2 + 
     3 m_\ini^2) m_W^2 - 2 m_W^4,
\end{align}
$\vartheta(x)$ is the Heaviside step function, and
\begin{eqnarray}
\mu_\ini^2 &=& \sqrt{m_\ini^4+m_d^4+m_W^4-2m_\ini^2 m_d^2 - 2 m_\ini^2 m_W^2 - 2 m_d^2 m_W^2} \,, \nonumber\\
\mu_\fin^2 &=& \sqrt{m_\fin^4+m_d^4+m_W^4-2m_\fin^2 m_d^2 - 2 m_\fin^2 m_W^2 - 2 m_d^2 m_W^2} \,.
\end{eqnarray}
It should be noted that $  \text{Im}[{\cal F}_{\ini \fin  , d}]$ is obtained by exchanging the masses $m_\ini$ and $m_\fin$ in $\text{Im}[{\cal F}_{\fin \ini   , d}]$ . We note the important feature of $\text{Im}[{\cal F}_{\fin \ini   , d}]\neq 0 $ being generated only in the branches where the particle mass conditions $m_\ini>m_W+m_d$ or $m_\fin>m_W+m_d$ is recovered. This important threshold mass condition required to generate kinetic $CP$ asymmetry at loop level is ameliorated further in Ref. \cite{Balaji:2019fxd}. We note that we keep the initial and final state quark masses general in the above discussion, however in the special case where the top quark decays into light flavour quarks, only the first bracketed terms in Eq.~\eqref{eq:imaginarypartFgamma} and Eq.~\eqref{eq:imaginarypartFg} respectively survive because the mass condition $m_t>m_W+m_d$ is satisfied. 

\subsection{Derivation of $CP$ asymmetry}
\label{section:CPasymmetry}
For Dirac particles, we state the $CP$ asymmetry between the initial and final state fermions as $u_\ini \to u_\fin \gamma_+$ and $\bar{u}_\ini \to \bar{u}_\fin \gamma_-$ and between $u_\ini \to u_\fin \gamma_-$ and $\bar{u}_\ini \to \bar{u}_\fin \gamma_+$, following similar notation to Ref. \cite{Balaji:2019fxd}. These can be written in terms of the photon polarisations (analogous replacements used for the gluon case) as  
\begin{eqnarray}
\Delta_{CP,+} = \frac{\Gamma(u_\ini \to u_\fin \gamma_+) - 
\Gamma(\bar{u}_\ini \to \bar{u}_\fin \gamma_-)}
{\Gamma(u_\ini \to u_\fin \gamma) + 
\Gamma(\bar{u}_\ini \to \bar{u}_\fin \gamma)} \,, \quad
\Delta_{CP,-} = \frac{\Gamma(u_\ini \to u_\fin \gamma_-) - 
\Gamma(\bar{u}_\ini \to \bar{u}_\fin \gamma_+)}
{\Gamma(u_\ini \to u_\fin \gamma) + 
\Gamma(\bar{u}_\ini \to \bar{u}_\fin \gamma)}\,,
\end{eqnarray}
and it then follows according to Ref. \cite{Balaji:2020fxd} that the $CP$ asymmetries can be written in terms of particle masses, CKM mixing and the loop functions $\mathcal{F}$ as 
\begin{eqnarray} 
\label{eq:CPparameters}
\Delta_{CP, +} &=& \frac{ - \sum_{\alpha,\beta}{\cal J}_{\alpha\beta}^{\ini \fin}
 {\rm Im}({\cal F}_{\ini \fin, \alpha} {\cal F}_{\ini \fin, \beta}^*) m_\fin^2 
 }{
 \sum_{\alpha,\beta}
{\cal R}_{\alpha\beta}^{\ini \fin}
\left[ {\rm Re}({\cal F}_{\fin \ini, \alpha} {\cal F}_{\fin \ini, \beta}^*) m_\ini^2 + {\rm Re}({\cal F}_{\ini \fin, \alpha} {\cal F}_{\ini \fin, \beta}^*) m_\fin^2 \right]
}\,, \nonumber\\
\Delta_{CP, -} &=& \frac{ - \sum_{\alpha,\beta}{\cal J}_{\alpha\beta}^{\ini \fin}
{\rm Im}({\cal F}_{\fin \ini, \alpha} {\cal F}_{\fin \ini, \beta}^*) m_\ini^2 
 }{
 \sum_{\alpha,\beta}
{\cal R}_{\alpha\beta}^{\ini \fin}
\left[ {\rm Re}({\cal F}_{\fin \ini, \alpha} {\cal F}_{\fin \ini, \beta}^*) m_\ini^2 + {\rm Re}({\cal F}_{\ini \fin, \alpha} {\cal F}_{\ini \fin, \beta}^*) m_\fin^2 \right]
} \,.
\end{eqnarray}
where $\alpha,\beta$ run for charged down-quark flavours $d,s,b$ and
\begin{eqnarray} \label{eq:J_and_R}
{\cal J}_{\alpha\beta}^{\ini \fin} = {\rm Im} ({V}_{ \ini \alpha } {V}_{\fin\alpha }^* {V}_{\ini \beta}^* {V}_{\fin \beta }) \,, &&
{\cal R}_{\alpha\beta}^{\ini \fin} = {\rm Re} ({V}_{\ini\alpha } {V}_{\fin\alpha }^* {V}_{\ini \beta }^* {V}_{\fin \beta }) \,.
\end{eqnarray}
Here the classic Jarlskog-like parameters ${\cal J}_{\alpha\beta}^{\ini \fin}$ are utilised to describe the $CP$ violation \cite{Jarlskog:1985ht, Wu:1985ea}. These parameters are invariant under any phase rotation of charged up and down-type quarks.
\section{Results}
\label{section:results}
\subsection{Branching ratios and decay widths}
\label{section:brsanddecaywidths}
In the SM, we may write the expression for the polarised radiative decay width in terms of functions denoted $A$ and $B$ for each channel as \cite{Balaji:2019fxd}
\begin{align}
\label{eq:polarisedwidths}
    \Gamma(t \rightarrow u_\beta \gamma_+)=\frac{1}{\pi}\left(\frac{m_t^2-m_u^2}{2m_t}\right)^3|A^\gamma-B^\gamma|^2\,,&& 
    \Gamma(t \rightarrow u_\beta \gamma_-)=\frac{1}{\pi}\left(\frac{m_t^2-m_u^2}{2m_t}\right)^3|A^\gamma+B^\gamma|^2,\nonumber\\
      \Gamma(t \rightarrow u_\beta g_+)=\frac{C_F}{\pi}\left(\frac{m_t^2-m_u^2}{2m_t}\right)^3|A^g-B^g|^2\,,&& 
    \Gamma(t \rightarrow u_\beta g_-)=\frac{C_F}{\pi}\left(\frac{m_t^2-m_u^2}{2m_t}\right)^3|A^g+B^g|^2.
\end{align}
Then it follows that the total unpolarised radiative width is given by summing the two polarisation channels and averaging over the two initial state spins so $\Gamma(t\rightarrow u_\beta \gamma)=\frac{1}{2}\left[\Gamma(t\rightarrow u_\beta \gamma_+)+\Gamma(t\rightarrow u_\beta \gamma_-)\right]$, which yields 
\begin{align}
\label{eq:unpolarisedwidths}
    \Gamma(t\rightarrow u_\beta\gamma)&=\frac{1}{\pi}\left(\frac{m_t^2-m_u^2}{2m_t}\right)^3\left(|A^\gamma|^2+|B^\gamma|^2\right)\,, &&
     \Gamma(t\rightarrow u_\beta g)&=\frac{C_F}{\pi}\left(\frac{m_t^2-m_u^2}{2m_t}\right)^3\left(|A^g|^2+|B^g|^2\right).
\end{align}
where $C_F=4/3$ is the standard colour factor \cite{AguilarSaavedra:2002ns}. We note that the usual Lorentz invariant amplitude can be separated into terms proportional and not proportional to $\gamma_5$ as
\begin{align}
\label{eq:matrixelement}
     \mathcal{M} (t \to u_\beta + \gamma)  = i \bar{u}(p_\beta) \sigma^{\mu \nu}(A^\gamma+B^\gamma \gamma_5)q_\nu u(p_t) \varepsilon^*_{\pm,\mu}(q) \,, 
\end{align}
by comparing coefficients between Eq. \eqref{eq:effective_vertex} and Eq. \eqref{eq:matrixelement}, it follows
\begin{align}
     A^\gamma = \frac{eG_F}{8\sqrt{2}\pi^2} V_{td} V_{u d}^{*}({\cal F}_{ut, d}^\gamma m_t + {\cal F}_{tu, d}^\gamma m_u)\,, && 
     B^\gamma = \frac{eG_F}{8\sqrt{2}\pi^2} V_{t d} V_{u d}^{*}({\cal F}_{ut, d}^\gamma m_t - {\cal F}_{tu, d}^\gamma m_u),
\end{align}
where $u=(u,c)$ depending on the final state and the above expressions must be summed over $d=(d,s,b)$ as shown in Eq. \eqref{eq:CPparameters} with each of their individual contributions. The corresponding parameters for gluon radiation, $A^g$ and $B^g$ , are obtained by simply performing the gauge coupling replacement $e\rightarrow g_s$ and ${\cal F}^\gamma \rightarrow {\cal F}^g$. We may also explicitly write the relations between the magnetic and electric transition dipole moments in terms of $A$ and $B$ as $f^M=-A^\gamma$ and $f^E=iB^\gamma$ \cite{Balaji:2019fxd}, the chromodynamic transition dipole moments are analogous except with the replacements $A^g$ and $B^g$ respectively.

The leading order SM top decay width is dominated by the tree level decay $t\rightarrow b W^+$ and given as \cite{AguilarSaavedra:2002ns}  
\begin{align}
    \Gamma(t \rightarrow b W^+)=\frac{g^2}{64\pi}|V_{tb}|^2 \frac{m_t^3}{m_W^2}\left(1-3 \frac{m_W^4}{m_t^4}+2 \frac{m_W^6}{m_t^6}\right).
\end{align}
We avoid using the next to leading order width as it makes a negligible difference numerically and our other calculations are performed at leading order. The branching ratios for the radiative processes are then simply given by
\begin{align}
\label{eq:branchingratio}
    {\cal B}(t\rightarrow u_\beta \gamma)=\frac{ \Gamma(t\rightarrow u_\beta\gamma)}{\Gamma(t \rightarrow b W^+)},
\end{align}
where the analogous replacement $\Gamma(t\rightarrow u_\beta g)$ is performed in the numerator when computing ${\cal B}(t\rightarrow u_\beta g) $.

\subsection{Numerical results and discussion}
\label{section:numericalresults}
We compute the branching ratios and $CP$ asymmetries according to Eq. \eqref{eq:branchingratio} and Eq. \eqref{eq:CPparameters} respectively. In this work, we use the standard parametrisation for the CKM matrix with angles $\theta_{12}=13.04\pm0.05^{\circ}$, $\theta_{13}=0.201\pm0.011^{\circ}$, $\theta_{23}=2.38\pm0.06^{\circ}$ and $\delta_{cp}=1.20\pm0.08$ \cite{Tanabashi:2018oca}. Additionally, we take the $b$-quark mass to be be the three loop $\overline{\text{MS}}$ scheme value evaluated at the top mass $m_b(m_t)=2.681\pm0.003$ \cite{Bednyakov:2016onn}. We take pole masses of $(m_t,m_c,m_u)=(173.21, 1.275,2.30\times10^{-3})$GeV for the external quarks. It should be noted that the running mass for the down-type quarks is not a fundamental parameter of the SM Lagrangian, but rather a product of the running Yukawa coupling $y_b=m_b/v$ and the Higgs vacuum expectation value $v$. Firstly, it is first of interest to directly calculate the central polarised widths which we obtain directly from Eq.~\eqref{eq:polarisedwidths} as

\begin{table}[ht!]  
    \centering
    \begin{tabular}{|c|c|c|c|} 
    \hline
     Decay Channel & Decay Width [GeV]   &  Decay Channel &  Decay Width [GeV] \\\hline
    $t\to u \gamma_+$& $ 2.714\times 10^{-21}$ & $t\to u g_+$ & $ 5.418\times 10^{-19}$ \\\hline
     $t\to u \gamma_-$&  $9.781  \times 10^{-16}$ &$t\to u g_-$ & $ 1.142 \times 10^{-13}$ \\\hline
     $t\to c \gamma_+$     & $1.520\times 10^{-18} $  & $t\to c g_+$    &$ 3.031\times 10^{-16} $ \\\hline
     $t\to c \gamma_-$     &  $1.364\times 10^{-13} $ & $t\to c g_-$   &$ 1.592\times10^{-11}$ \\\hline
    \end{tabular}
    \caption{Results for the polarised decay widths for the radiative channels $t\rightarrow u\gamma$, $t\rightarrow c\gamma$, $t\rightarrow u g$ and $t\rightarrow c g$.}
    \label{table:polarisedwidthstable}
\end{table}

The total unpolarised branching ratios can then be computed from Eq.~\eqref{eq:unpolarisedwidths}, which are shown in Table \ref{table:resultstable} and are approximately one order of magnitude smaller compared to the ones quoted in Ref. \cite{Eilam:1990zc}. This is expected as they used the internal $b$-quark pole mass in their calculation ($m_b = 5$ GeV is assumed). In the more recent Ref.~\cite{AguilarSaavedra:2002ns}, they compute 
\begin{align}
    \mathcal{B}(t\rightarrow u\gamma)\simeq 3.7\times10^{-16}\,, &&
     \mathcal{B}(t\rightarrow c\gamma)\simeq 4.6\times 10^{-14},\nonumber \\
         \mathcal{B}(t\rightarrow ug)\simeq 3.7\times10^{-14}\,, &&
     \mathcal{B}(t\rightarrow cg)\simeq 4.6\times 10^{-12},
\end{align}
 which is comparable to those shown in Table \ref{table:resultstable}, the marginal differences observed are well within the one sigma uncertainties they quote and can be attributed to the fact that they use a now superseded running mass for the $b$-quark of $m_b(m_t) = 2.74\pm0.17$ GeV as well as an external line $c$-quark mass of $m_c=1.5$ GeV. As previously noted in the same work, the uncertainty in the top quark mass does not affect the results shown, since the partial widths of $t \rightarrow c\gamma$, $t \rightarrow c g$ are proportional to $m_t^3$, it follows that the leading dependence on $m_t$ gets cancelled when branching ratios and $CP$ asymmetries are computed, meaning the uncertainty in $m_t$ has a negligible effect on the final result.

In Ref. \cite{AguilarSaavedra:2002ns}, they also provide an order of magnitude estimate for the $CP$ asymmetries
\begin{align}
    \Delta_{{CP},-}(t\rightarrow c\gamma)\sim -5\times10^{-6}\,, &&
     \Delta_{{CP},-}(t\rightarrow c g)\sim -6\times10^{-6},
\end{align}
 in the SM case\footnote{We note that the $CP$ asymmetries are denoted $a_\gamma$ and $a_g$ in Ref. \cite{AguilarSaavedra:2002ns}, corresponds to $\Delta_{{CP}}=\Delta_{{CP},+} + \Delta_{{CP},-}$. In this work $\Delta_{{CP},+}\ll \Delta_{{CP},-}$ and so $\Delta_{{CP}}\simeq \Delta_{{CP},-}$}. This is about a factor of two smaller than the result we compute in Table \ref{table:resultstable}. This is an unsurprising discrepancy as the result shown in this work includes all of the kinetic terms, appropriate quark running masses and current CKM parameters. Here we see that the ratio for branching fractions and the 
 $CP$-asymmetries can be approximated $\frac{\mathcal{B}(t\rightarrow 
 c\gamma(g))}{\mathcal{B}(t\rightarrow u\gamma(g))}\simeq  
 \left(\frac{|V_{cb}|}{|V_{ub}|}\right)^2$, $\frac{\Delta_{CP,-}(t\rightarrow 
 c\gamma(g))}{\Delta_{CP,-}(t\rightarrow u\gamma(g))}\simeq 
 -\left(\frac{|V_{ub}|}{|V_{cb}|}\right)^2$ $\frac{\Delta_{CP,+}(t\rightarrow 
 c\gamma(g))}{\Delta_{CP,+}(t\rightarrow u\gamma(g))}\simeq 
 \frac{|V_{cb}|}{|V_{ub}|}\frac{m_c}{m_u}$ while the hierarchy $\Delta_{CP,+}\ll \Delta_{CP,-}$ is a direct consequence of angular momentum conservation and the fact that the weak interaction is parity violating. 
\begin{table}[ht!]  
    \centering
    \begin{tabular}{|c|c|c|c|} 
    \hline
     Decay Channel & Branching Ratio   &  $\Delta_{CP,+} \qquad $ &  $\Delta_{CP,-} \qquad $ \\\hline
    $t\to u \gamma$& $ (3.262 \pm 0.341)\times 10^{-16}$ & $ -( 7.142 \pm 0.668 )\times 10^{-14} $ & $ (1.612  \pm 0.151)\times 10^{-3}$ \\\hline
     $t\to c \gamma$&  $(4.550  \pm 0.234) \times 10^{-14}$ &$ -(6.232  \pm0.605 )\times10^{-10} $ & $ -(1.150  \pm 0.112 ) \times 10^{-5}$ \\\hline
     $t\to u g$     & $(3.810  \pm 0.340)\times 10^{-14} $  & $ -(4.521  \pm 0.424 )\times10^{-14} $   &$ ( 1.617 \pm 0.152)\times 10^{-3} $ \\\hline
     $t\to c g$     &  $(5.310  \pm 0.271)\times 10^{-12} $ & $ -(6.245 \pm0.605)\times10^{-10} $   &$ -(1.153  \pm 0.112 )\times10^{-5}$ \\\hline
    \end{tabular}
    \caption{Results for the branching ratio and $CP$ asymmetries for the radiative channels $t\rightarrow u\gamma$, $t\rightarrow c\gamma$, $t\rightarrow u g$ and $t\rightarrow c g$. The quoted uncertainty is propagated from the one sigma CKM angle uncertainties and running bottom quark mass at the top quark mass scale using the $\overline{\text{MS}}$ scheme.}
    \label{table:resultstable}
\end{table}
\section{Application to selected new physics models}
\label{section:BSMmodels}
We do not focus on beyond the SM physics scenarios in this work, however there are numerous potential applications of the results shown in this paper to beyond the SM theories. The most direct of these is likely the aforementioned extension of the SM via VLQs. This is motivated, namely by a recent more precise evaluation of $V_{ud}$ and $V_{us}$, which places the unitarity condition of the first row in the CKM matrix $|V_{ud}|^2+|V_{us}|^2 +|V_{ub}|^2 =
0.99798\pm0.00038$ at a deviation more than $4\sigma$ from unity \cite{Belfatto:2019swo, Seng:2018yzq}. Furthermore, a mild excess in the overall Higgs signal strength appears at about $2\sigma$ above the standard model (SM) prediction \cite{Sirunyan:2018koj}. Additionally, there is the long-lasting discrepancy in the forward-backward asymmetry $\mathcal{A}^b_{FB}$ in $Z\rightarrow b \overline{b}$ at LEP \cite{Tanabashi:2018oca}. There have been models motivated by explaining the above three anomalies via extension of the SM quark sector via down-type VLQs which alleviate the tension among these datasets such as the one shown in Ref. \cite{Cheung:2020vqm}. 

There are also direct searches for the down-type VLQs at the LHC \cite{Aaboud:2018pii, Sirunyan:2018qau, Sirunyan:2019sza}. Inclusion of these down-type quarks $b'$ and $b''$ realise improved agreement to data compared to the SM \cite{Cheung:2020vqm}. The results shown in Section \ref{section:kineticterms} in conjunction with Section \ref{section:CPasymmetry} can be used to predict polarised photons observables resulting from the $CP$ asymmetries for processes such as $b'\rightarrow d_\beta \gamma$ and $b''\rightarrow d_\beta \gamma$. It should be noted that these VLQs are experimentally favoured over previously studied fourth generation models such as in Ref. \cite{Eilam:2009hz} due to precision Higgs measurements at the LHC \cite{Erler:2010sk}. The main addition to the results shown in this work for a complete description of these decays would be the inclusion of FCNC diagrams with $Z$, $h$ and unphysical scalar $\chi$ bosons appearing in the penguin diagrams. However, it should be noted that these amplitudes share similar Lorentz structure to the results shown in this paper. Hence, this class of models represent a relatively straightforward extension. We plan to show this explicitly in a future work. Experimental interest in such models is high and there has been many detailed searches performed for these down-type VLQs at the LHC \cite{Aaboud:2018pii, Sirunyan:2018qau, Sirunyan:2019sza, Sirunyan:2017pks, Aaboud:2018ifs}. Similarly, in Ref. \cite{Alok:2015iha}, the inclusion of new vector iso-singlet up-type quarks is discussed in detail with a $4\times 3$ CKM matrix. ATLAS searches have also already been conducted to try and find these new up-type quarks, which are often referred to as $t'$ or $T$ in the literature \cite{Chatrchyan:2012fp,Chatrchyan:2013uxa}. 

Additionally, the mass hierarchy between the up-type and down-type quarks observed in nature motivates  consideration of models with two complex $SU(2)_L$ doublet scalar fields which comprise the 2HDM. In the so called type III 2HDM both doublets simultaneously give masses to all quark types. In these 2HDM variants, it has been shown that $\mathcal{B}(t\rightarrow c\gamma)$ can reach about $ 10^{-8}$ \cite{Luke:1993cy}, $10^{-6}$ \cite{Diaz:2001vj,Gaitan-Lozano:2014nka} and recently it has even been suggested that parameter regions exist where it can be enhanced to about $ 10^{-5}$. The dominant contributions for the rare radiative top decay $t \rightarrow c\gamma$ at one-loop in 2HDM come from neutral and charged Higgs bosons running in the loop analogous to the third diagram in Figure \ref{figure:FeynmanDiagramsgamma} but with the $W$ bosons replaced with the charged Higgs $H^+$ and the second diagram where the unphysical scalar $\chi$ is replaced with the physical SM-like Higgs $h$. Therefore, it is clear that the result for the $CP$ asymmetry shown in this paper can easily exported for use in the 2HDM as well. We note that the previous focus in the literature of these rare decays has primarily been on photon radiation rather than gluon radiation. The latter of which, we have studied in this work and would be expected to have a much larger branching fraction albeit a less experimentally clean probe of new physics in hadron colliders due to large quantum chromodynamics (QCD) backgrounds.

\section{Conclusion}
The rare radiative flavour changing loop level top decays $t\to c\gamma$, $t\to cg$, $t\to u\gamma$ and $t\to ug$ branching ratios and corresponding $CP$ asymmetries are computed in full detail. These signatures exist due to imaginary components of the loop functions and the CKM matrix and provide a potentially clean probe of new physics or further validation of the SM. A full analytical formulation for the $CP$ asymmetry resulting from the loop functions as well as a revised numerical computation of the SM branching fractions is provided. The branching fractions are comparable to the values quoted in the literature while the $CP$ asymmetry is computed to a higher degree of precision and is about a factor of two larger than the previously stated order of magnitude estimates \cite{AguilarSaavedra:2002ns}. These rare radiative processes are suppressed in the SM by the GIM mechanism, however, the kinetic terms and loop functions presented can easily be adapted for use with minimal modification in extensions of the SM via vector-like quarks or in Two-Higgs-Doublet models. These extensions can enhance the same channels of interest by many orders of magnitude relative to the SM, even reaching branching ratios up to $10^{-5}$ or higher, due to the presence of an extended CKM matrix, FCNC at tree level or new scalar field content respectively. Several of these extensions have been studied in detail recently and comprise an active area of research since they can provide improved global fits to several recent flavour physics measurements. Studying the phenomenology of radiative decays produced in these beyond the SM models by application of the formulae detailed is intended to be performed as a future work. 
\section*{Acknowledgements} 

SB would like to thank Maura Ramirez-Quezada and Ye-Ling Zhou for many useful discussions and helpful feedback. SB would also like to extend thanks to C\'eline B\oe hm and Kevin Varvell for suggestions that improved the clarity of the final text.
\clearpage
\bibliographystyle{JHEP} 
\bibliography{refs}

\providecommand{\href}[2]{#2}\begingroup\raggedright\begin{thebibliography}{10}

\bibitem{Beneke:2000hk}
M.~Beneke et~al., \emph{{Top quark physics}},  in \emph{{Workshop on Standard
  Model Physics (and more) at the LHC (First Plenary Meeting)}}, pp.~419--529,
  3, 2000.
\newblock \href{http://arxiv.org/abs/hep-ph/0003033}{{\tt hep-ph/0003033}}.

\bibitem{Balaji:2019fxd}
S.~Balaji, M.~Ramirez-Quezada and Y.-L. Zhou, \emph{{CP violation and circular
  polarisation in neutrino radiative decay}},
  \href{http://dx.doi.org/10.1007/JHEP04(2020)178}{\emph{JHEP} {\bf 04} (2020)
  178}, [\href{http://arxiv.org/abs/1910.08558}{{\tt 1910.08558}}].

\bibitem{Balaji:2020fxd}
S.~Balaji, M.~Ramirez-Quezada and Y.-L. Zhou, \emph{{CP violation in the
  neutrino dipole moment}},  \href{http://arxiv.org/abs/2008.12795}{{\tt
  2008.12795}}.

\bibitem{AguilarSaavedra:2002ns}
J.~Aguilar-Saavedra and B.~Nobre, \emph{{Rare top decays t ---> c gamma, t --->
  cg and CKM unitarity}},
  \href{http://dx.doi.org/10.1016/S0370-2693(02)03230-6}{\emph{Phys. Lett. B}
  {\bf 553} (2003) 251--260}, [\href{http://arxiv.org/abs/hep-ph/0210360}{{\tt
  hep-ph/0210360}}].

\bibitem{Eilam:1990zc}
G.~Eilam, J.~Hewett and A.~Soni, \emph{{Rare decays of the top quark in the
  standard and two Higgs doublet models}},
  \href{http://dx.doi.org/10.1103/PhysRevD.44.1473}{\emph{Phys. Rev. D} {\bf
  44} (1991) 1473--1484}.

\bibitem{Atwood:1996vj}
D.~Atwood, L.~Reina and A.~Soni, \emph{{Phenomenology of two Higgs doublet
  models with flavor changing neutral currents}},
  \href{http://dx.doi.org/10.1103/PhysRevD.55.3156}{\emph{Phys. Rev. D} {\bf
  55} (1997) 3156--3176}, [\href{http://arxiv.org/abs/hep-ph/9609279}{{\tt
  hep-ph/9609279}}].

\bibitem{Grzadkowski:1990sm}
B.~Grzadkowski, J.~Gunion and P.~Krawczyk, \emph{{Neutral current flavor
  changing decays for the Z boson and the top quark in two Higgs doublet
  models}}, \href{http://dx.doi.org/10.1016/0370-2693(91)90931-F}{\emph{Phys.
  Lett. B} {\bf 268} (1991) 106--111}.

\bibitem{Luke:1993cy}
M.~E. Luke and M.~J. Savage, \emph{{Flavor changing neutral currents in the
  Higgs sector and rare top decays}},
  \href{http://dx.doi.org/10.1016/0370-2693(93)90238-D}{\emph{Phys. Lett. B}
  {\bf 307} (1993) 387--393}, [\href{http://arxiv.org/abs/hep-ph/9303249}{{\tt
  hep-ph/9303249}}].

\bibitem{Tanabashi:2018oca}
{\scshape ParticleDataGroup} collaboration, M.~Tanabashi et~al., \emph{{Review
  of Particle Physics}},
  \href{http://dx.doi.org/10.1103/PhysRevD.98.030001}{\emph{Phys. Rev.} {\bf
  D98} (2018) 030001}.

\bibitem{AguilarSaavedra:2002kr}
J.~Aguilar-Saavedra, \emph{{Effects of mixing with quark singlets}},
  \href{http://dx.doi.org/10.1103/PhysRevD.69.099901}{\emph{Phys. Rev. D} {\bf
  67} (2003) 035003}, [\href{http://arxiv.org/abs/hep-ph/0210112}{{\tt
  hep-ph/0210112}}].

\bibitem{Alok:2015iha}
A.~K. Alok, S.~Banerjee, D.~Kumar, S.~U. Sankar and D.~London,
  \emph{{New-physics signals of a model with a vector-singlet up-type quark}},
  \href{http://dx.doi.org/10.1103/PhysRevD.92.013002}{\emph{Phys. Rev. D} {\bf
  92} (2015) 013002}, [\href{http://arxiv.org/abs/1504.00517}{{\tt
  1504.00517}}].

\bibitem{Alok:2015vvk}
A.~K. Alok, S.~Banerjee, D.~Kumar, S.~U. Sankar and D.~London,
  \emph{{New-physics signals of a model with an isosinglet vector-like t'
  quark}}, \href{http://dx.doi.org/10.22323/1.234.0579}{\emph{PoS} {\bf
  EPS-HEP2015} (2015) 579}.

\bibitem{Alok:2014yua}
A.~K. Alok, S.~Banerjee, D.~Kumar and S.~Uma~Sankar, \emph{{Flavor signatures
  of isosinglet vector-like down quark model}},
  \href{http://dx.doi.org/10.1016/j.nuclphysb.2016.03.012}{\emph{Nucl. Phys. B}
  {\bf 906} (2016) 321--341}, [\href{http://arxiv.org/abs/1402.1023}{{\tt
  1402.1023}}].

\bibitem{Aguilar-Saavedra:2013qpa}
J.~Aguilar-Saavedra, R.~Benbrik, S.~Heinemeyer and M.~Pérez-Victoria,
  \emph{{Handbook of vectorlike quarks: Mixing and single production}},
  \href{http://dx.doi.org/10.1103/PhysRevD.88.094010}{\emph{Phys. Rev. D} {\bf
  88} (2013) 094010}, [\href{http://arxiv.org/abs/1306.0572}{{\tt 1306.0572}}].

\bibitem{Boehm:2003hm}
C.~Boehm and P.~Fayet, \emph{{Scalar dark matter candidates}},
  \href{http://dx.doi.org/10.1016/j.nuclphysb.2004.01.015}{\emph{Nucl. Phys. B}
  {\bf 683} (2004) 219--263}, [\href{http://arxiv.org/abs/hep-ph/0305261}{{\tt
  hep-ph/0305261}}].

\bibitem{Gaitan:2015hga}
R.~Gaitán, J.~H. Montes~de Oca, E.~A. Garcés and R.~Martinez, \emph{{Rare top
  decay $t \rightarrow c \gamma$ with flavor changing neutral scalar
  interactions in two Higgs doublet model}},
  \href{http://dx.doi.org/10.1103/PhysRevD.94.094038}{\emph{Phys. Rev. D} {\bf
  94} (2016) 094038}, [\href{http://arxiv.org/abs/1503.04391}{{\tt
  1503.04391}}].

\bibitem{Dedes:2014asa}
A.~Dedes, M.~Paraskevas, J.~Rosiek, K.~Suxho and K.~Tamvakis, \emph{{Rare
  Top-quark Decays to Higgs boson in MSSM}},
  \href{http://dx.doi.org/10.1007/JHEP11(2014)137}{\emph{JHEP} {\bf 11} (2014)
  137}, [\href{http://arxiv.org/abs/1409.6546}{{\tt 1409.6546}}].

\bibitem{Hill:1994hp}
C.~T. Hill, \emph{{Topcolor assisted technicolor}},
  \href{http://dx.doi.org/10.1016/0370-2693(94)01660-5}{\emph{Phys. Lett. B}
  {\bf 345} (1995) 483--489}, [\href{http://arxiv.org/abs/hep-ph/9411426}{{\tt
  hep-ph/9411426}}].

\bibitem{Arhrib:2005nx}
A.~Arhrib, \emph{{Top and Higgs flavor changing neutral couplings in two Higgs
  doublets model}},
  \href{http://dx.doi.org/10.1103/PhysRevD.72.075016}{\emph{Phys. Rev. D} {\bf
  72} (2005) 075016}, [\href{http://arxiv.org/abs/hep-ph/0510107}{{\tt
  hep-ph/0510107}}].

\bibitem{DiazCruz:1989ub}
J.~Diaz-Cruz, R.~Martinez, M.~Perez and A.~Rosado, \emph{{Flavor Changing
  Radiative Decay of Thf T Quark}},
  \href{http://dx.doi.org/10.1103/PhysRevD.41.891}{\emph{Phys. Rev. D} {\bf 41}
  (1990) 891--894}.

\bibitem{Atwood:1995ud}
D.~Atwood, L.~Reina and A.~Soni, \emph{{Probing flavor changing top - charm -
  scalar interactions in $e^{+} e^{-}$ collisions}},
  \href{http://dx.doi.org/10.1103/PhysRevD.53.1199}{\emph{Phys. Rev. D} {\bf
  53} (1996) 1199--1201}, [\href{http://arxiv.org/abs/hep-ph/9506243}{{\tt
  hep-ph/9506243}}].

\bibitem{Atwood:1995ej}
D.~Atwood, L.~Reina and A.~Soni, \emph{{Flavor changing neutral scalar currents
  at $\mu^{+} \mu^{-}$ colliders}},
  \href{http://dx.doi.org/10.1103/PhysRevLett.75.3800}{\emph{Phys. Rev. Lett.}
  {\bf 75} (1995) 3800--3803}, [\href{http://arxiv.org/abs/hep-ph/9507416}{{\tt
  hep-ph/9507416}}].

\bibitem{Balaji:2018zna}
S.~Balaji, R.~Foot and M.~A. Schmidt, \emph{{Chiral SU(4) explanation of the
  $b\to s$ anomalies}},
  \href{http://dx.doi.org/10.1103/PhysRevD.99.015029}{\emph{Phys. Rev. D} {\bf
  99} (2019) 015029}, [\href{http://arxiv.org/abs/1809.07562}{{\tt
  1809.07562}}].

\bibitem{Balaji:2019kwe}
S.~Balaji and M.~A. Schmidt, \emph{{Unified SU(4) theory for the $R_{D^{(*)}}$
  and $R_{K^{(*)}}$ anomalies}},
  \href{http://dx.doi.org/10.1103/PhysRevD.101.015026}{\emph{Phys. Rev. D} {\bf
  101} (2020) 015026}, [\href{http://arxiv.org/abs/1911.08873}{{\tt
  1911.08873}}].

\bibitem{Boehm:2017nrl}
C.~B\oe~hm, C.~Degrande, O.~Mattelaer and A.~C. Vincent, \emph{{Circular
  polarisation: a new probe of dark matter and neutrinos in the sky}},
  \href{http://dx.doi.org/10.1088/1475-7516/2017/05/043}{\emph{JCAP} {\bf 05}
  (2017) 043}, [\href{http://arxiv.org/abs/1701.02754}{{\tt 1701.02754}}].

\bibitem{Jarlskog:1985ht}
C.~Jarlskog, \emph{{Commutator of the Quark Mass Matrices in the Standard
  Electroweak Model and a Measure of Maximal CP Violation}},
  \href{http://dx.doi.org/10.1103/PhysRevLett.55.1039}{\emph{Phys. Rev. Lett.}
  {\bf 55} (1985) 1039}.

\bibitem{Wu:1985ea}
D.-d. Wu, \emph{{The Rephasing Invariants and CP}},
  \href{http://dx.doi.org/10.1103/PhysRevD.33.860}{\emph{Phys. Rev. D} {\bf 33}
  (1986) 860}.

\bibitem{Bednyakov:2016onn}
A.~Bednyakov, B.~Kniehl, A.~Pikelner and O.~Veretin, \emph{{On the $b$-quark
  running mass in QCD and the SM}},
  \href{http://dx.doi.org/10.1016/j.nuclphysb.2017.01.004}{\emph{Nucl. Phys. B}
  {\bf 916} (2017) 463--483}, [\href{http://arxiv.org/abs/1612.00660}{{\tt
  1612.00660}}].

\bibitem{Belfatto:2019swo}
B.~Belfatto, R.~Beradze and Z.~Berezhiani, \emph{{The CKM unitarity problem: A
  trace of new physics at the TeV scale?}},
  \href{http://dx.doi.org/10.1140/epjc/s10052-020-7691-6}{\emph{Eur. Phys. J.
  C} {\bf 80} (2020) 149}, [\href{http://arxiv.org/abs/1906.02714}{{\tt
  1906.02714}}].

\bibitem{Seng:2018yzq}
C.-Y. Seng, M.~Gorchtein, H.~H. Patel and M.~J. Ramsey-Musolf, \emph{{Reduced
  Hadronic Uncertainty in the Determination of $V_{ud}$}},
  \href{http://dx.doi.org/10.1103/PhysRevLett.121.241804}{\emph{Phys. Rev.
  Lett.} {\bf 121} (2018) 241804}, [\href{http://arxiv.org/abs/1807.10197}{{\tt
  1807.10197}}].

\bibitem{Sirunyan:2018koj}
{\scshape CMS} collaboration, A.~M. Sirunyan et~al., \emph{{Combined
  measurements of Higgs boson couplings in proton\textendash{}proton collisions
  at $\sqrt{s}=13\,\text {Te}\text {V} $}},
  \href{http://dx.doi.org/10.1140/epjc/s10052-019-6909-y}{\emph{Eur. Phys. J.
  C} {\bf 79} (2019) 421}, [\href{http://arxiv.org/abs/1809.10733}{{\tt
  1809.10733}}].

\bibitem{Cheung:2020vqm}
K.~Cheung, W.-Y. Keung, C.-T. Lu and P.-Y. Tseng, \emph{{Vector-like Quark
  Interpretation for the CKM Unitarity Violation, Excess in Higgs Signal
  Strength, and Bottom Quark Forward-Backward Asymmetry}},
  \href{http://dx.doi.org/10.1007/JHEP05(2020)117}{\emph{JHEP} {\bf 05} (2020)
  117}, [\href{http://arxiv.org/abs/2001.02853}{{\tt 2001.02853}}].

\bibitem{Aaboud:2018pii}
{\scshape ATLAS} collaboration, M.~Aaboud et~al., \emph{{Combination of the
  searches for pair-produced vector-like partners of the third-generation
  quarks at $\sqrt{s} =$ 13 TeV with the ATLAS detector}},
  \href{http://dx.doi.org/10.1103/PhysRevLett.121.211801}{\emph{Phys. Rev.
  Lett.} {\bf 121} (2018) 211801}, [\href{http://arxiv.org/abs/1808.02343}{{\tt
  1808.02343}}].

\bibitem{Sirunyan:2018qau}
{\scshape CMS} collaboration, A.~M. Sirunyan et~al., \emph{{Search for
  vector-like quarks in events with two oppositely charged leptons and jets in
  proton-proton collisions at $\sqrt{s} =$ 13 TeV}},
  \href{http://dx.doi.org/10.1140/epjc/s10052-019-6855-8}{\emph{Eur. Phys. J.
  C} {\bf 79} (2019) 364}, [\href{http://arxiv.org/abs/1812.09768}{{\tt
  1812.09768}}].

\bibitem{Sirunyan:2019sza}
{\scshape CMS} collaboration, A.~M. Sirunyan et~al., \emph{{Search for pair
  production of vectorlike quarks in the fully hadronic final state}},
  \href{http://dx.doi.org/10.1103/PhysRevD.100.072001}{\emph{Phys. Rev. D} {\bf
  100} (2019) 072001}, [\href{http://arxiv.org/abs/1906.11903}{{\tt
  1906.11903}}].

\bibitem{Eilam:2009hz}
G.~Eilam, B.~Melic and J.~Trampetic, \emph{{CP violation and the 4th
  generation}}, \href{http://dx.doi.org/10.1103/PhysRevD.80.116003}{\emph{Phys.
  Rev. D} {\bf 80} (2009) 116003}, [\href{http://arxiv.org/abs/0909.3227}{{\tt
  0909.3227}}].

\bibitem{Erler:2010sk}
J.~Erler and P.~Langacker, \emph{{Precision Constraints on Extra Fermion
  Generations}},
  \href{http://dx.doi.org/10.1103/PhysRevLett.105.031801}{\emph{Phys. Rev.
  Lett.} {\bf 105} (2010) 031801}, [\href{http://arxiv.org/abs/1003.3211}{{\tt
  1003.3211}}].

\bibitem{Sirunyan:2017pks}
{\scshape CMS} collaboration, A.~M. Sirunyan et~al., \emph{{Search for pair
  production of vector-like quarks in the bW$\overline{\mathrm{b}}$W channel
  from proton-proton collisions at $\sqrt{s} =$ 13 TeV}},
  \href{http://dx.doi.org/10.1016/j.physletb.2018.01.077}{\emph{Phys. Lett. B}
  {\bf 779} (2018) 82--106}, [\href{http://arxiv.org/abs/1710.01539}{{\tt
  1710.01539}}].

\bibitem{Aaboud:2018ifs}
{\scshape ATLAS} collaboration, M.~Aaboud et~al., \emph{{Search for single
  production of vector-like quarks decaying into $Wb$ in $pp$ collisions at
  $\sqrt{s} = 13$ TeV with the ATLAS detector}},
  \href{http://dx.doi.org/10.1007/JHEP05(2019)164}{\emph{JHEP} {\bf 05} (2019)
  164}, [\href{http://arxiv.org/abs/1812.07343}{{\tt 1812.07343}}].

\bibitem{Chatrchyan:2012fp}
{\scshape CMS} collaboration, S.~Chatrchyan et~al., \emph{{Combined Search for
  the Quarks of a Sequential Fourth Generation}},
  \href{http://dx.doi.org/10.1103/PhysRevD.86.112003}{\emph{Phys. Rev. D} {\bf
  86} (2012) 112003}, [\href{http://arxiv.org/abs/1209.1062}{{\tt 1209.1062}}].

\bibitem{Chatrchyan:2013uxa}
{\scshape CMS} collaboration, S.~Chatrchyan et~al., \emph{{Inclusive Search for
  a Vector-Like T Quark with Charge $\frac{2}{3}$ in pp Collisions at
  $\sqrt{s}$ = 8 TeV}},
  \href{http://dx.doi.org/10.1016/j.physletb.2014.01.006}{\emph{Phys. Lett. B}
  {\bf 729} (2014) 149--171}, [\href{http://arxiv.org/abs/1311.7667}{{\tt
  1311.7667}}].

\bibitem{Diaz:2001vj}
R.~A. Diaz, R.~Martinez and J.~Alexis~Rodriguez, \emph{{The Rare decay $t \to c
  \gamma$ in the general 2 HDM type III}},
  \href{http://arxiv.org/abs/hep-ph/0103307}{{\tt hep-ph/0103307}}.

\bibitem{Gaitan-Lozano:2014nka}
R.~Gaitan-Lozano, R.~Martinez and J.~H.~M. de~Oca, \emph{{Rare top decay
  $t\rightarrow c\gamma$ in general THDM-III}},
  \href{http://arxiv.org/abs/1407.3318}{{\tt 1407.3318}}.

\end{thebibliography}\endgroup
\end{document}